%% file: main.tex
\documentclass[sigconf]{aamas} 
\usepackage{balance} %
\usepackage{amsmath}
\usepackage{amsfonts}
\usepackage{bm}
\usepackage{amsthm}
\usepackage{algorithm}
\usepackage{algorithmic}
\usepackage{nicefrac}
\usepackage{booktabs}
\usepackage{cleveref}
\usepackage{xcolor}

\newcommand{\papertitle}{Learning Competitive Equilibria in Noisy Combinatorial Markets}

\input{include/commands}
\title[\papertitle]{\papertitle}

\author{Enrique Areyan Viqueira}
\affiliation{
  \institution{Brown University}
  \city{Providence, RI}}
\email{eareyan@brown.edu}

\author{Cyrus Cousins}
\affiliation{
  \institution{Brown University}
  \city{Providence, RI}}
\email{cyrus\_cousins@brown.edu}

\author{Amy Greenwald}
\affiliation{
  \institution{Brown University}
  \city{Providence, RI}}
\email{amy_greenwald@brown.edu}

\input{abstract}


\newcommand{\BibTeX}{\rm B\kern-.05em{\sc i\kern-.025em b}\kern-.08em\TeX}

\newtheorem{theorem}{Theorem}
\newtheorem{definition}{Definition}
\newtheorem{lemma}{Lemma}
\newtheorem{observation}{Observation}

\makeatletter
\def\@copyrightspace{\relax}
\makeatother
\setcopyright{none}

\begin{document}

\pagestyle{fancy}
\fancyhead{}

\maketitle 

\input{intro}

\input{related}

\input{model}
\input{learn}

\input{algos}
\input{prune}
\input{expts}

\input{value_models}

\input{conc}

\input{ack}

\balance
\bibliographystyle{ACM-Reference-Format}
\bibliography{include/bib}

\clearpage
\input{appendix}

\end{document}

%% file: include/commands.tex
\newcommand{\mydef}[1]{\emph{#1}}

\newcommand{\alternative}[1]{#1'}

\newcommand{\SetOfGoods}{G}
\newcommand{\SetOfGoodsAlt}{\alternative{\SetOfGoods}}
\newcommand{\NumberOfGoods}{m}
\newcommand{\Good}{j}
\newcommand{\Bundle}{S}
\newcommand{\BundleAlt}{T}

\newcommand{\Buyer}{i}

\newcommand{\SetOfConsumers}{N}
\newcommand{\SetOfConsumersAlt}{\alternative{\SetOfConsumers}}
\newcommand{\NumberOfConsumers}{n}
\newcommand{\Consumer}{i}
\newcommand{\ConsumerAlt}{k}
\newcommand{\Valuation}{v}
\newcommand{\ValuationFunction}[1]{\Valuation_{#1}}
\newcommand{\ValuationFunctionAlt}[1]{\alternative{\Valuation_{#1}}}

\newcommand{\Market}{M}
\newcommand{\MarketAlt}{\alternative{\Market}}

\newcommand{\SubMarket}{\Market_{-(\Consumer, \Bundle)}}
\newcommand{\SubMarketHat}{\hat{\Market}_{-(\Consumer, \Bundle)}}
\newcommand{\SubMarketAlt}{\alternative{\Market}_{-(\Consumer, \Bundle)}}

\newcommand{\Allocation}{\mathcal{S}}
\newcommand{\AllocationAlt}{\mathcal{U}}
\newcommand{\SetOfFeasibleAllocations}{\mathcal{F}}
\newcommand{\Welfare}{w}
\newcommand{\WelfareUpBound}[2]{w_{(#1,#2)}^{\diamond}}
\newcommand{\Pricing}{\mathcal{P}}
\newcommand{\Price}{p}
\newcommand{\PriceFunction}{P}

\newcommand{\UM}{UM}
\newcommand{\RM}{RM}
\newcommand{\CESet}{\mathcal{CE}}

\newcommand{\SetOfConditions}{\mathcal{X}}
\newcommand{\Condition}{x}
\newcommand{\ConditionDistribution}{\mathcal{D}}
\newcommand{\ExpectedValuationFunction}[1]{\Valuation_{#1}}
\newcommand{\EmpiricalValuationFunction}[1]{\hat{\Valuation}_{#1}}
\newcommand{\ConditionalMarket}[1]{\Market_{#1}}
\newcommand{\ExpectedMarket}[1]{\Market_{#1}}
\newcommand{\EmpiricalMarket}[1]{\hat{\Market}_{#1}}
\newcommand{\VectorOfConditions}{\bm{x}}
\newcommand{\NumberOfSamples}{t}
\newcommand{\SampleIndex}{l}
\newcommand{\IndexSet}{\mathcal{I}}
\newcommand{\ValuationRange}{c}

\newcommand{\PruningIterationIndex}{k}
\newcommand{\SamplingSchedule}{\bm{\NumberOfSamples}}
\newcommand{\SamplingScheduleIndex}[1]{\NumberOfSamples_{#1}}
\newcommand{\DeltaSchedule}{\bm{\delta}}
\newcommand{\DeltaScheduleIndex}[1]{\delta_{#1}}
\newcommand{\PruningSchedule}{\bm{\pi}}
\newcommand{\PruningScheduleIndex}[1]{\pi_{#1}}

\newcommand{\EA}{\ensuremath{\operatorname{EA}}}
\newcommand{\EAP}{\ensuremath{\operatorname{EAP}}}

\newcommand{\norm}[1]{\left\lVert#1\right\rVert}
\newcommand{\Expectation}{\mathbf{E}}
\newcommand{\Probability}{\mathbb{P}}

\newcommand{\UnitDemandMatrixRV}{\mathcal{V}}
\newcommand{\UnitDemandMatrix}{\mathbf{V}}
\newcommand{\UMLoss}{\emph{UM-Loss}}

\renewcommand{\epsilon}{\varepsilon}

\newcommand{\ifequals}[3]{\ifthenelse{\equal{#1}{#2}}{#3}{}}

%% file: abstract.tex
\begin{abstract}
  We present a methodology to robustly estimate the competitive equilibria (CE) of combinatorial markets under the assumption that buyers do not know their precise valuations for bundles of goods, but instead can only provide noisy estimates. We first show tight lower- and upper-bounds on the buyers' utility loss, and hence the set of CE, given a uniform approximation of one market by another. We then develop a learning framework for our setup, and present two probably-approximately-correct algorithms for learning CE, i.e., producing uniform approximations that preserve CE, with finite-sample guarantees. The first is a baseline that uses Hoeffding's inequality to produce a uniform approximation of buyers' valuations with high probability. The second leverages 
  a connection between the first welfare theorem of economics and uniform approximations to adaptively prune value queries when it determines that they are provably not part of a CE. We experiment with our algorithms and find that the pruning algorithm achieves better estimates than the baseline with far fewer samples.
\end{abstract}

%% file: intro.tex
\section{Introduction}

Combinatorial Markets (CMs) are a class of markets in which buyers are interested in acquiring bundles of goods, and their values for these bundles can be arbitrary. 
Real-world examples of CMs include:
    spectrum auctions~\cite{cramton2002spectrum}
    allocation of landing and take-off slots at airports~\cite{ball2006auctions};
	internet ad placement~\cite{edelman2007internet}; and
	procurement of bus routes~\cite{cantillon2006auctioning}.
An outcome of a CM is an assignment of bundles to buyers together with prices for the goods. %
A competitive equilibrium (CE) is an outcome of particular interest in CMs and other well-studied economic models~\cite{bikhchandani1997competitive,leon2003elements}. In a CE, buyers are utility-maximizing (i.e., they maximize their utilities among all feasible allocations at the posted prices) and the seller maximizes its revenue (again, over all allocations at the posted prices).

While CEs are a static equilibrium concept, they can sometimes arise as the outcome of a dynamic price adjustment process (e.g.,~\cite{cheung2020tatonnement}). In such a process, prices might be adjusted by an imaginary Walrasian auctioneer, who poses demand queries to buyers: i.e., asks them their demands at given prices. Similarly, we imagine that prices in a CM are set by a market maker, who poses value queries to buyers: i.e., asks them their values on select bundles.

One of the defining features of CMs is that they afford buyers the flexibility to express complex preferences,
which in turn has the potential to increase market efficiency.
However, the extensive expressivity of these markets presents challenges for both the market maker and the buyers.
With an exponential number of bundles in general, it is infeasible for a buyer to evaluate them all.
We thus present a model of noisy buyer valuations:
e.g., buyers might use approximate or heuristic methods
to obtain value estimates~\cite{fujishima1999taming}.
In turn, the market maker chooses an outcome in the face of uncertainty about the buyers' valuations.
We call the objects of study in this work \mydef{noisy combinatorial markets} (NCM) to emphasize that buyers do not have direct access to their values for bundles, but instead can only noisily estimate them.

In this work, we formulate a mathematical model of NCMs. 
Our goal is then to design learning algorithms with rigorous finite-sample guarantees that approximate the competitive equilibria of NCMs.
Our first result is to show tight lower- and upper-bounds on the set of CE, given uniform approximations of buyers' valuations.
We then present two learning algorithms.
The first one---Elicitation Algorithm; \EA---serves as a baseline.
It uses Hoeffding's inequality~\cite{hoeffding1994probability} to produce said uniform approximations.
Our second algorithm---Elicitation Algorithm with Pruning; \EAP---leverages 
the first welfare theorem of economics
to adaptively prune value queries 
when it determines that they are provably not part of a CE.

After establishing the correctness of our algorithms, we evaluate their empirical performance
using both synthetic unit-demand valuations and two spectrum auction value models. The former are a class of valuations central to the literature on economics and computation~\cite{lehmann2006combinatorial}, for which there are efficient algorithms to compute CE~\cite{gul1999walrasian}. 
In the spectrum auction value models, the buyers' valuations are characterized by complements, which complicate the questions of existence and computability of CE.
In all three models,
we measure the average quality of learned CE via our algorithms, compared to the CE of the corresponding certain market (i.e., here, ``certain'' means lacking uncertainty), as a function of the number of samples.
We find that \EAP{} often yields better error guarantees than \EA{} using far fewer samples,
because it successfully prunes buyers' valuations (i.e., it ceases querying for values on bundles of goods that a CE provably does not comprise), even without any \emph{a priori\/} knowledge of the market's combinatorial structure. 

As the market size grows, an interesting tradeoff arises between computational and sample efficiency. 
To prune a value query and retain rigorous guarantees on the quality of the learned CE, we must solve a welfare-maximizing problem whose complexity grows with the market's size. Consequently, at each iteration of \EAP, for each value query, we are faced with a choice. Either solve said welfare-maximizing problem and potentially prune the value query (thereby saving on future samples), or defer attempts to prune the value query, until more is known about the market.
To combat this situation,
we show that an upper bound on the optimal welfare's value (rather than the precise value) suffices
to obtain rigorous guarantees on the learned CE's quality. 
Such upper bounds can be found easily, by solving a relaxation of the welfare-maximization problem.
Reminiscent of designing admissible heuristics in classical search problems, 
this methodology applies to any combinatorial market, but at the same time allows for the application of domain-dependent knowledge to compute these upper bounds, when available.
Empirically, we show that a computationally cheap relaxation of the welfare-maximization problem yields substantial sample and computational savings in a large market.

%% file: related.tex
\paragraph{Related Work}
The idea for this paper stemmed from the work on abstraction in Fisher markets by~\citeauthor{kroer2019computing}~\cite{kroer2019computing}.
There, the authors tackle the problem of computing equilibria in large markets by creating an abstraction of the market, computing equilibria in the abstraction, and lifting those equilibria back to the original market.
Likewise, we develop a pruning criterion which in effect builds an abstraction of any CM, where then compute a CE, which is provably also an approximate CE in the original market.

The mathematical formalism we adopt follows that of \citeauthor{areyan2020improved}~\cite{areyan2020improved}.
There, the authors propose a mathematical framework for empirical game-theoretic analysis~\cite{wellman2006methods}, and algorithms that learn the Nash equilibria of simulation-based games~\cite{vorobeychik2008stochastic,vorobeychik2010probabilistic}.
In this paper, we extend this methodology to market equilibria, and provide analogous results in the case of CMs.
Whereas intuitively, a basic pruning criterion for games is arguably more straightforward---simply prune dominated strategies---the challenge in this work was to discover a pruning criterion that would likewise prune valuations that are provably not part of a CE.

\citeauthor{jha2020learning}~\cite{jha2020learning} have also tackled the problem of learning CE in CM.%
\footnote{The market structure they investigate is not identical to the structure studied here. Thus, at present, our results are not directly comparable.}
Whereas our approach is to accurately learn only those components of the buyers' valuations that determine a CE (up to PAC guarantees), their approach bypasses the learning of agent preferences altogether, going straight for learning a solution concept, such as a CE.
It is an open 
question as to whether one approach dominates the other, in the context of noisy CMs.

Another related line of research is concerned with learning valuation functions from data~\cite{balcan2012learning,balcan2011learning,lahaie2004applying}.
In contrast, our work is concerned with learning buyers' valuations only in so much as it facilitates learning CE.
Indeed, our main conclusion is that CE often can be learned from just a subset of the buyers' valuations.

There is also a long line of work on preference elicitation in combinatorial auctions (e.g.,~\cite{conen2001preference}), where an auctioneer aims to pose value queries in an intelligent order so as to minimize the computational burden on the bidders, while still clearing the auction.

Finally, our pruning criterion relies on a novel application of the first welfare theorem of economics.
While prior work
has connected economic theory with algorithmic complexity~\cite{roughgarden2015prices}, this work connects economic theory with statistical learning theory.

%% file: model.tex
\section{Model}
\label{sec:model}

We write $\mathbb{X}_+$ to denote the set of positive values in a numerical set $\mathbb{X}$ including zero. Given an integer $k \in \mathbb{Z}$, we write $[k]$ to denote the first $k$ integers, inclusive: i.e., $[k] = \{ 1, 2, \ldots, k \}$. Given a finite set of integers $Z \subset \mathbb{Z}$, we write $2^Z$ to denote the power set of $Z$.

A \mydef{combinatorial market} is defined by a set of goods and a set of buyers.
We denote the set of goods by $\SetOfGoods = [\NumberOfGoods]$,
and the set of buyers by $\SetOfConsumers = [\NumberOfConsumers]$.
We index an arbitrary good by $\Good \in \SetOfGoods$, 
and an arbitrary buyer by $\Consumer \in \SetOfConsumers$.
A \mydef{bundle} of goods is a set of goods $\Bundle \subseteq \SetOfGoods$.
Each buyer $\Consumer$ is characterized by their preferences over bundles, represented as a valuation function $\ValuationFunction{\Consumer}: 2^\SetOfGoods \mapsto \mathbb{R}_+$, where $\ValuationFunction{\Consumer} (\Bundle) \in \mathbb{R}_+$ is buyer $\Consumer$'s value for bundle $\Bundle$.
We assume valuations are normalized so that $\ValuationFunction{\Consumer} (\emptyset) = 0$, for all $\Consumer \in \SetOfConsumers$.
Using this notation, a combinatorial market---market, hereafter---is a tuple $\Market = (\SetOfGoods, \SetOfConsumers, \{\ValuationFunction{\Consumer}\}_{\Consumer \in \SetOfConsumers})$. 

Given a market $\Market$, an \mydef{allocation} $\Allocation = (\Bundle_1, \ldots, \Bundle_\NumberOfConsumers)$ denotes an assignment of goods to buyers, where $\Bundle_\Consumer \subseteq \SetOfGoods$ is the bundle assigned to buyer $\Consumer$. 
We consider only feasible allocations.
An allocation $\Allocation$ is \mydef{feasible} if $\Bundle_\Consumer \cap \Bundle_\ConsumerAlt = \emptyset$ for all $\Consumer, \ConsumerAlt \in \SetOfConsumers$ such that $\Consumer \neq \ConsumerAlt$.
We denote the set of all feasible allocations of market $\Market$ by $\SetOfFeasibleAllocations(\Market)$.
The \mydef{welfare} of allocation $\Allocation$ is defined as $\Welfare(\Allocation) = \sum_{\Consumer \in \SetOfConsumers} \ValuationFunction{\Consumer} (\Bundle_\Consumer)$.
A welfare-maximizing allocation $\Allocation^*$ is a feasible allocation that yields maximum welfare among all feasible allocations, i.e., $\Allocation^* \in \arg\max_{\Allocation \in \SetOfFeasibleAllocations(\Market)} \Welfare (\Market)$. We denote by $\Welfare^*(\Market)$ the welfare of any welfare-maximizing allocation $\Allocation^*$, i.e., $\Welfare^*(\Market) = \Welfare(\Allocation^*) = \sum_{\Consumer \in \SetOfConsumers} \Valuation_\Consumer(\Bundle^*_\Consumer)$.

A pricing profile $\Pricing = (\PriceFunction_1, \ldots, \PriceFunction_{\NumberOfConsumers})$ is a vector of $\NumberOfConsumers$ pricing functions, one function  $\PriceFunction_\Consumer: 2^\SetOfGoods \mapsto \mathbb{R}_+$ for each buyer, each mapping bundles to prices, $\PriceFunction_\Consumer(\Bundle) \in \mathbb{R}_+$. The seller's revenue of allocation $\Allocation$ given a pricing $\Pricing$ is $\sum_{\Consumer \in \SetOfConsumers} \PriceFunction_\Consumer(\Bundle_\Consumer)$. 
We refer to pair $(\Allocation, \Pricing)$ as a \mydef{market outcome}---outcome, for short. Given an outcome, buyer $\Consumer$'s utility is difference between its attained value and its payment, $\ValuationFunction{\Consumer}(\Bundle_\Consumer) - \PriceFunction_\Consumer(\Bundle_\Consumer)$, and the seller's utility is equal to its revenue. 

In this paper, we are interested in approximations of one market by another. We now define a mathematical framework in which to formalize such approximations. In what follows, whenever we decorate a market $\Market$, e.g., $\MarketAlt$, what we mean is that we decorate each of its components: i.e., $\MarketAlt = (\SetOfGoodsAlt, \SetOfConsumersAlt, \{\ValuationFunctionAlt{\Consumer}\}_{\Consumer \in \SetOfConsumersAlt})$.

It will be convenient to refer to a subset of buyer--bundle pairs.
We use the notation $\IndexSet \subseteq \SetOfConsumers \times 2^\SetOfGoods$ for this purpose. 

Markets $\Market$ and $\MarketAlt$ are \mydef{compatible} if $\SetOfGoods = \SetOfGoodsAlt$ and $\SetOfConsumers = \SetOfConsumersAlt$. 
{Whenever a market $\Market$ is compatible with a market $\MarketAlt$, an outcome of $\Market$ is also an outcome of $\MarketAlt$.}
Given two compatible markets $\Market$ and $\MarketAlt$, we measure the difference between them at $\IndexSet$ as
$\norm{\Market - \MarketAlt}_{\IndexSet} = 
        \max_{(\Consumer, \Bundle) \in \IndexSet} 
            | \ValuationFunction{\Consumer} (\Bundle) - \ValuationFunctionAlt{\Consumer} (\Bundle) |$.
When $\IndexSet = \SetOfConsumers \times 2^\SetOfGoods$, this difference is precisely the infinity norm.
Given $\epsilon > 0$, $\Market$ and $\MarketAlt$ are called \emph{$\epsilon$-approximations} of one another if 
$\norm{\Market - \MarketAlt}_\infty \le \epsilon$.

The solution concept of interest in this paper is competitive equilibrium\footnote{A competitive equilibrium is always guaranteed to exists~\cite{bikhchandani2002package}.}. A competitive equilibrium consists of two conditions: the utility-maximization (\UM) condition and the revenue-maximization (\RM) condition. \UM{} ensures that the allocation maximizes buyers' utilities given the pricings, while \RM{} ensures that the seller maximizes its utility. Together, both conditions constitute an equilibrium of the market, i.e., an outcome where no agent has an incentive to deviate by, for example, relinquishing its allocation. We now formalize this solution concept, followed by its relaxation, central when working with approximate markets.

\begin{definition}[Competitive Equilibrium]
Given a market $\Market$, an outcome $(\Allocation, \Pricing)$ is a \mydef{competitive equilibrium} (CE) if:

\begin{itemize}
    \item[(\UM)]
    $\forall \Consumer \in \SetOfConsumers, \BundleAlt \subseteq \SetOfGoods : 
        \ValuationFunction{\Consumer} (\Bundle_\Consumer) 
            - 
        \PriceFunction_\Consumer (\Bundle_\Consumer)
            \ge
        \ValuationFunction{\Consumer} (\BundleAlt) 
            - 
        \PriceFunction_\Consumer (\BundleAlt)$   
    
    \item[(\RM)]
    $\forall \Allocation' \in \SetOfFeasibleAllocations(\Market): 
        \sum_{\Consumer \in \SetOfConsumers} \PriceFunction_\Consumer(\Bundle_\Consumer) 
        \ge 
        \sum_{\Consumer \in \SetOfConsumers} \PriceFunction_\Consumer(\Bundle_\Consumer') 
        $
    
\end{itemize}
\end{definition}

\begin{definition}[Approximate Competitive Equilibria]
Let $\epsilon > 0$. 
An outcome $(\Allocation, \Pricing)$ is a $\epsilon$-competitive equilibrium ($\epsilon$-CE) if it is a CE in which \UM{} holds up to $\epsilon$:

    $\epsilon$-(\UM)
    $\forall \Consumer \in \SetOfConsumers, \BundleAlt \subseteq \SetOfGoods : 
        \ValuationFunction{\Consumer} (\Bundle_\Consumer) 
            - 
        \PriceFunction_\Consumer (\Bundle_\Consumer)
            + 
            \epsilon
            \ge
        \ValuationFunction{\Consumer} (\BundleAlt) 
            - 
        \PriceFunction_\Consumer (\BundleAlt)$ 
\end{definition}

For $\alpha \ge 0$, we denote by $\CESet_\alpha(\Market)$ the set of all $\alpha$-approximate CE of $\Market$, i.e., $\CESet_\alpha (\Market) = \{ (\Allocation, \Pricing) : (\Allocation, \Pricing)$ is a $\alpha$-approximate CE of $\Market \}$. Note that $\CESet_0 (\Market)$ is the set of (exact) CE of market $\Market$, which we denote $\CESet (\Market)$.

\begin{theorem}[Competitive Equilibrium Approximation]
\label{thm:CEApprox}
Let $\epsilon > 0$.
If $\Market$ and $\MarketAlt$ are compatible markets such that $\norm{\Market - \MarketAlt}_\infty \le \epsilon$, then $\CESet (\Market) \subseteq \CESet_{2 \epsilon} (\MarketAlt) \subseteq \CESet_{4 \epsilon} (\Market)$.
\end{theorem}

\begin{proof}
We prove that:
$\CESet_{\alpha} (\Market) \subseteq \CESet_{\alpha + 2\epsilon} (\MarketAlt)$, for $\alpha \ge 0$. This result then implies $\CESet (\Market) \subseteq \CESet_{2 \epsilon} (\MarketAlt)$ when $\alpha = 0$; likewise, it (symmetrically) implies $\CESet_{2 \epsilon} (\MarketAlt) \subseteq \CESet_{4 \epsilon} (\Market)$ when $\alpha = 2 \epsilon$.

Let $\Market$ and $\MarketAlt$ be compatible markets s.t.\ $\norm{\Market - \MarketAlt}_\infty \le \epsilon$. Suppose $(\Allocation, \Pricing)$ is a $\alpha$-competitive equilibrium of $\Market$. Our task is to show that $(\Allocation, \Pricing)$, interpreted as an outcome of $\MarketAlt$, is a $(\alpha + 2\epsilon)$-competitive equilibrium of $\MarketAlt$.

First, note that the \RM{} condition is immediately satisfied, because $\Allocation$ and $\Pricing$ do not change when interpreting $(\Allocation, \Pricing)$ as an outcome of $\MarketAlt$. Thus, we need only show that the approximation holds for the \UM{} condition:
\begin{align}
    \ValuationFunctionAlt{\Consumer} (\Bundle_\Consumer) 
    - \PriceFunction_\Consumer (\Bundle_\Consumer) 
        & \ge 
    \ValuationFunction{\Consumer} (\Bundle_\Consumer) 
    - \PriceFunction_\Consumer (\Bundle_\Consumer) - \epsilon, & \forall \Consumer, \Bundle_\Consumer \label{thm1:eps-close-1}\\
        & \ge 
    \ValuationFunction{\Consumer} (\BundleAlt) 
    - \PriceFunction_\Consumer (\BundleAlt) - \alpha - \epsilon,& \forall \BundleAlt\subseteq\SetOfGoods \label{thm1:approx-ce}\\
        & \ge
    \ValuationFunctionAlt{\Consumer} (\BundleAlt) 
    - \PriceFunction_\Consumer (\BundleAlt) - \alpha - 2\epsilon,& \forall \BundleAlt\subseteq\SetOfGoods \label{thm1:eps-close-2}
\end{align}

\noindent
where (\ref{thm1:eps-close-1}) and (\ref{thm1:eps-close-2}) follow because $\norm{\Market - \MarketAlt}_\infty \le \epsilon$, and (\ref{thm1:approx-ce}) follows because $(\Allocation, \Pricing)$ is a $\alpha$-approximate CE of $\Market$.
\end{proof}

%% file: learn.tex
\section{Learning Methodology}
\label{sec:learn}

We now present a formalism in which to model noisy combinatorial markets.
Intuitively, a noisy market is one in which buyers' valuations over bundles are not known precisely;
rather, only noisy samples are available.

\begin{definition}[Conditional Combinatorial Markets]
\label{def:ConditionalCM}
A \mydef{conditional comb. market} $\ConditionalMarket{\SetOfConditions} = (\SetOfConditions, \SetOfGoods, \SetOfConsumers, \{\ValuationFunction{\Consumer}\}_{\Consumer \in \SetOfConsumers})$ consists of a set of conditions $\SetOfConditions$, a set of goods $\SetOfGoods$, a set of buyers $\SetOfConsumers$, and a set of conditional valuation functions $\{\ValuationFunction{\Consumer}\}_{\Consumer \in \SetOfConsumers}$, where $\ValuationFunction{\Consumer} : 2^\SetOfGoods \times \SetOfConditions \mapsto \mathbb{R}_+$. Given a condition $\Condition \in \SetOfConditions$, the value $\ValuationFunction{\Consumer} (\Bundle, \Condition)$ is $\Consumer$'s value for bundle $\Bundle \subseteq \SetOfGoods$. 
\end{definition}

\begin{definition}[Expected Combinatorial Market]
\label{def:EmpiricalCM}
Let $\ConditionalMarket{\SetOfConditions} = (\SetOfConditions, $ $\SetOfGoods, \SetOfConsumers, \{\ValuationFunction{\Consumer}\}_{\Consumer \in \SetOfConsumers})$ be a conditional combinatorial market and let $\ConditionDistribution$ be a distribution over $\SetOfConditions$.
For all $\Consumer \in \SetOfConsumers$, define the expected valuation function $\ExpectedValuationFunction{\Consumer} : 2^\SetOfGoods \mapsto \mathbb{R}_+$ by $\ExpectedValuationFunction{\Consumer} (\Bundle, \ConditionDistribution) = \Expectation_{\Condition \sim \ConditionDistribution}[\ValuationFunction{\Consumer} (\Bundle, \Condition)]$, and the corresponding expected combinatorial market as $\ExpectedMarket{\ConditionDistribution} = (\SetOfGoods, \SetOfConsumers, \{\ExpectedValuationFunction{\Consumer}\}_{\Consumer \in \SetOfConsumers})$.
\end{definition}

The goal of this work is to design algorithms that learn the approximate CE of expected combinatorial markets.
We will learn their equilibria given access only to their empirical counterparts, which we define next.

\begin{definition}[Empirical Combinatorial Market]
\label{def:ExpectedCM}
Let $\ConditionalMarket{\SetOfConditions} = (\SetOfConditions, $ $\SetOfGoods, \SetOfConsumers, \{\ValuationFunction{\Consumer}\}_{\Consumer \in \SetOfConsumers})$ be a conditional combinatorial market and let $\ConditionDistribution$ be a distribution over $\SetOfConditions$.
Denote by $\VectorOfConditions = (\Condition_1, \ldots, \Condition_\NumberOfSamples) \sim \ConditionDistribution$ a vector of $\NumberOfSamples$
samples drawn from $\SetOfConditions$ according to distribution $\ConditionDistribution$.
For all $\Consumer \in \SetOfConsumers$, we define the empirical valuation function $\EmpiricalValuationFunction{\Consumer} : 2^\SetOfGoods \mapsto \mathbb{R}_+$ by $\smash{\EmpiricalValuationFunction{\Consumer} (\Bundle) = \frac{1}{\NumberOfSamples} \sum_{\SampleIndex = 1}^\NumberOfSamples \ValuationFunction{\Consumer} (\Bundle, \Condition_\SampleIndex)}$, and the corresponding empirical combinatorial market as $\EmpiricalMarket{\VectorOfConditions} = (\SetOfGoods, \SetOfConsumers, \{\EmpiricalValuationFunction{\Consumer}\}_{\Consumer \in \SetOfConsumers})$.
\end{definition}

\begin{observation}[Learnability]
Let $\ConditionalMarket{\SetOfConditions}$ be a conditional combinatorial market and let $\ConditionDistribution$ be a distribution over $\SetOfConditions$.
Let $\ExpectedMarket{\ConditionDistribution}$ and $\EmpiricalMarket{\VectorOfConditions}$ be the corresponding expected and empirical combinatorial markets. If, for some $\epsilon, \delta > 0$, it holds that $\small\smash{\Probability\left(\norm{\ExpectedMarket{\ConditionDistribution} - \EmpiricalMarket{\VectorOfConditions}} \le \epsilon\right) \ge 1 -\delta}$, then the competitive equilibria of $\ExpectedMarket{\ConditionDistribution}$ are learnable: i.e, any competitive equilibrium of $\ExpectedMarket{\ConditionDistribution}$ is a $2 \epsilon$-competitive equilibrium of $\EmpiricalMarket{\VectorOfConditions}$ with probability at least $1 - \delta$.
\end{observation}

\Cref{thm:CEApprox} implies that CE are approximable to within any desired $\epsilon > 0$ guarantee. The following lemma shows  we only need a finitely many samples to learn them to within any $\delta > 0$ probability.

\begin{lemma}[Finite-Sample Bounds for Expected Combinatorial Markets via Hoeffding's Inequality]
\label{lemma:hoeffding}
Let $\ConditionalMarket{\SetOfConditions}$ be a conditional combinatorial market, $\ConditionDistribution$ a distribution over $\SetOfConditions$, and $\IndexSet \subseteq \SetOfConsumers \times 2^\SetOfGoods$ an index set. Suppose that for all $\Condition \in \SetOfConditions$ and $(\Consumer, \Bundle) \in \IndexSet$, it holds that $\ValuationFunction{\Consumer} (\Bundle, \Condition) \in [0, \ValuationRange]$ where $\ValuationRange \in \mathbb{R}_+$. Then, with probability at least $1 - \delta$ over samples $\VectorOfConditions = (\Condition_1, \ldots, \Condition_\NumberOfSamples) \sim \ConditionDistribution$, it holds that
$\smash{\norm{\ExpectedMarket{\ConditionDistribution} - \EmpiricalMarket{\VectorOfConditions}}_{\IndexSet}} \le 
    \ValuationRange
    \sqrt{\nicefrac{\ln(\nicefrac{2|\IndexSet|}{\delta})}{2\NumberOfSamples}}$, where $\delta > 0$. (Proof in the Appendix)
\end{lemma}

Hoeffding's inequality is a convenient and simple bound, where only knowledge of the range of values is required.
However, the union bound can be inefficient in large combinatorial markets.
This shortcoming can be addressed via uniform convergence bounds and Rademacher averages \citep{bartlett2002rademacher,areyan2019learning,koltchinskii2001rademacher}.
Furthermore, sharper empirical variance sensitive bounds have been shown to improve sample complexity in learning the Nash equilibria of black-box games \citep{areyan2020improved}.
In particular, to obtain a confidence interval of radius $\varepsilon$ in a combinatorial market with index set $\IndexSet = \SetOfConsumers \times 2^\SetOfGoods$, Hoeffding's inequality requires $t \in \smash{\mathcal{O}(\nicefrac{\ValuationRange^2|\SetOfGoods|}{\varepsilon^{2}}\ln\nicefrac{|\SetOfConsumers|}{\delta})}$ samples.
Uniform convergence bounds can improve the $|\SetOfGoods|$ term arising from the \emph{union bound}, and variance-sensitive bounds can largely replace dependence on $\ValuationRange^{2}$ with \emph{variances}.
Nonetheless, even without these augmentations, our methods are statistically efficient in $|\SetOfGoods|$, requiring only \emph{polynomial} sample complexity to learn \emph{exponentially} large combinatorial markets.

%% file: algos.tex
\subsection{Baseline Algorithm}
\label{sec:baseline}

\EA{} (\Cref{alg:EA}) is a preference elicitation algorithm for combinatorial markets.
The algorithm places value queries, but is only assumed to elicit noisy values for bundles.
The following guarantee follows immediately from Lemma~\ref{lemma:hoeffding}.

\begin{theorem}[Elicitation Algorithm Guarantees] 
\label{thm:EAGuarantees}
Let $\ConditionalMarket{\SetOfConditions}$ be a conditional market, $\ConditionDistribution$ be a distribution over $\SetOfConditions$, $\IndexSet$ an index set, 
$\NumberOfSamples \in \mathbb{N}_{>0}$ a number of samples, 
$\delta > 0$, and $\ValuationRange \in \mathbb{R}_+$.
Suppose that for all $\Condition \in \SetOfConditions$ and $(\Consumer, \Bundle) \in \IndexSet$, it holds that $\ValuationFunction{\Consumer} (\Bundle, \Condition) \in [0, \ValuationRange]$. 
If~\EA{} outputs $(\{ \EmpiricalValuationFunction{\Consumer} \}_{(\Consumer, \Bundle) \in \IndexSet}, \hat{\epsilon})$ on input $(\ConditionalMarket{\SetOfConditions}, \ConditionDistribution, \IndexSet, \NumberOfSamples, \delta, \ValuationRange)$, then, with probability at least $1 - \delta$, it holds that 
$\smash{\norm{\ExpectedMarket{\ConditionDistribution} - \EmpiricalMarket{\VectorOfConditions}}_{\IndexSet}} \le 
        \ValuationRange
        \sqrt{\nicefrac{\ln(\nicefrac{2|\IndexSet|}{\delta})}{2\NumberOfSamples}}$.
\end{theorem}

\begin{proof}
The result follows from \Cref{lemma:hoeffding}.
\end{proof}

%% file: prune.tex
\subsection{Pruning Algorithm}
\label{sec:pruning}

\EA{} elicits buyers' valuations for all bundles,
but in certain situations, some buyer valuations are not relevant for computing a CE---although bounds on all of them are necessary to guarantee strong bounds on the set of CE (\Cref{thm:CEApprox}). 
For example, in a first-price auction for one good, it is enough to accurately learn the highest bid, but is not necessary to accurately learn all other bids, if it is known that they are lower than the highest.
Since our goal is to learn CE, we present \EAP{} (Algorithm~\ref{alg:EAP}), an algorithm that does not sample uniformly, but instead adaptively decides which value queries to prune so that, with provable guarantees, \EAP's estimated market satisfies the conditions of \Cref{thm:CEApprox}.

\input{algos/eap}
\input{algos/ea}

\EAP{} (\Cref{alg:EAP}) takes as input 
a sampling schedule $\SamplingSchedule$, 
a failure probability schedule $\DeltaSchedule$, 
and a pruning budget schedule $\PruningSchedule$. 
The sampling schedule $\SamplingSchedule$ is a sequence of $|\SamplingSchedule|$ strictly decreasing integers $\SamplingScheduleIndex{1} > \SamplingScheduleIndex{2} > \cdots > \SamplingScheduleIndex{|\SamplingSchedule|}$, where $\SamplingScheduleIndex{\PruningIterationIndex}$ is the total number of samples to take for each $(\Consumer, \Bundle)$ pair during \EAP's $\PruningIterationIndex$-th iteration. 
The failure probability schedule $\DeltaSchedule$ is a sequence of the same length as $\SamplingSchedule$, where $\DeltaScheduleIndex{\PruningIterationIndex} \in (0, 1)$ is the $\PruningIterationIndex$-th iteration's failure probability and $\sum_{\PruningIterationIndex} \DeltaScheduleIndex{\PruningIterationIndex} \in (0, 1)$ is the total failure probability.
The pruning budget schedule $\PruningSchedule$ is a sequence of integers also of the same length as $\SamplingSchedule$, where $\PruningScheduleIndex{\PruningIterationIndex}$ is the maximum number of $(\Consumer, \Bundle)$ pruning candidate pairs.
The algorithm progressively elicits buyers' valuations via repeated calls to \EA.
However, between calls to \EA, \EAP{} searches for value queries that are provably not part of a CE; the size of this search is dictated by the pruning schedule. All such queries (i.e., buyer--bundle pairs) then cease to be part of the index set with which \EA{} is called in future iterations.

In what follows, we prove several intermediate results, which enable us to prove the main result of this section, \Cref{thm:EAP}, which establishes \EAP's correctness.
Specifically, the market learned by \EAP---with potentially different numbers of samples for different $(\Consumer, \Bundle)$ pairs---is enough to provably recover any CE of the underlying market.

\begin{lemma}[Optimal Welfare Approximations]
\label{lemma:opt-welfare}
Let $\Market$ and $\MarketAlt$ be compatible markets such that they $\epsilon$-approximate one another.
Then $|\Welfare^*(\Market) - \Welfare^*(\MarketAlt)| \le \epsilon \NumberOfConsumers$.
\end{lemma}

\begin{proof}
Let $\Allocation^*$ be a welfare-maximizing allocation for $\Market$ and $\AllocationAlt^*$ be a welfare-maximizing allocation for $\MarketAlt$.
Let $\Welfare^*(\Market)$ be the maximum achievable welfare in market $\Market$.
Then, 
\input{equations/opt_welfare}
\noindent
The first inequality follows from the optimality of $\Allocation^*$ in $\Market$, and the second from
the $\epsilon$-approximation assumption.
Likewise, $\Welfare^*(\MarketAlt) \ge  \Welfare^*(\Market) - \epsilon \NumberOfConsumers$, so the result holds.
\end{proof}

The key to this work was the discovery of a pruning criterion that removes $(\Consumer, \Bundle)$ pairs from consideration if they are provably not part of any CE.
Our check relies on computing the welfare of the market without the pair: i.e., in submarkets.

\begin{definition}
Given a market $\Market$ and buyer--bundle pair $(\Consumer, \Bundle)$, the \mydef{$(\Consumer, \Bundle)$-submarket} of $\Market$, denoted by $\SubMarket$, is the market obtained by removing all goods in $\Bundle$ and buyer $\Consumer$ from market $\Market$. That is, $\SubMarket =  (\SetOfGoods \setminus \Bundle, 
    \SetOfConsumers \setminus \{\Consumer\}, \{\ValuationFunction{\ConsumerAlt}\}_{\ConsumerAlt \in \SetOfConsumers \setminus \{\Consumer\}})$.
\end{definition}

\begin{lemma}[Pruning Criteria]
\label{lemma:pruning}
Let $\Market$ and $\MarketAlt$ be compatible markets such that $\norm{\Market - \MarketAlt}_\infty \le \epsilon$.
In addition, let $(\Consumer, \Bundle)$ be a buyer, bundle pair, and $\SubMarketAlt$ be the $(\Consumer, \Bundle)$-submarket of $\MarketAlt$. Finally, let $\WelfareUpBound{\Consumer}{\Bundle} \in \mathbb{R}_+$ upper bound $\Welfare^*(\SubMarketAlt)$, i.e., $\Welfare^*(\SubMarketAlt) \le \WelfareUpBound{\Consumer}{\Bundle}$.
If the following pruning criterion holds, 
then $\Bundle$ is not allocated to $\Consumer$ in any welfare-maximizing allocation of $\Market$:
\begin{equation}
    \label{eq:pruning}
    \ValuationFunctionAlt{\Consumer} (\Bundle) +
        \WelfareUpBound{\Consumer}{\Bundle} + 2 \epsilon \NumberOfConsumers
        < \Welfare^*(\MarketAlt) \enspace .
\end{equation}
\end{lemma}

\begin{proof}

Let $\Allocation^*, \AllocationAlt^*,$ and $\AllocationAlt^*_{-(\Consumer, \Bundle)}$ be welfare-maximizing allocations of markets $\Market, \MarketAlt,$ and $\SubMarketAlt$, respectively.
Then, 
\begin{align}
    \Welfare^*(\Market) 
        & \ge \Welfare^*(\MarketAlt) - \epsilon \NumberOfConsumers \label{lemma-pruning:lemma-welfare} \\
        & > \ValuationFunctionAlt{\Consumer} (\Bundle) +
        \WelfareUpBound{\Consumer}{\Bundle} + \epsilon \NumberOfConsumers \\
        & \ge \ValuationFunctionAlt{\Consumer} (\Bundle) +
        \Welfare^*(\SubMarketAlt) + \epsilon \NumberOfConsumers \\
        & \ge \ValuationFunction{\Consumer} (\Bundle) - \epsilon + \Welfare^*(\SubMarket)  - \epsilon(\NumberOfConsumers - 1) + \epsilon \NumberOfConsumers \\
        & = \ValuationFunction{\Consumer} (\Bundle) + \Welfare^*(\SubMarket)
\end{align}
The first inequality follows from \Cref{lemma:opt-welfare}.
The second follows from Equation (\ref{eq:pruning}) and 
the third because $\WelfareUpBound{\Consumer}{\Bundle}$ is an upper bound of $\Welfare^*(\SubMarketAlt)$.
The fourth inequality follows from the assumption that $\norm{\Market - \MarketAlt}_\infty \le \epsilon$, and by \Cref{lemma:opt-welfare} applied to submarket $\SubMarket$. 
Therefore, the allocation where $\Consumer$ gets $\Bundle$ cannot be welfare-maximizing in market $\Market$. 
\end{proof}

\Cref{lemma:pruning} provides a family of pruning criteria parameterized by the upper bound $\smash{\WelfareUpBound{\Consumer}{\Bundle}}$. 
The closer $\smash{\WelfareUpBound{\Consumer}{\Bundle}}$ is to $\smash{\Welfare^*(\SubMarketAlt)}$, the sharper the pruning criterion, with the best pruning criterion being $\smash{\WelfareUpBound{\Consumer}{\Bundle} = \Welfare^*(\SubMarketAlt)}$.
However, solving for $\smash{\Welfare^*(\SubMarketAlt)}$ exactly can easily become a bottleneck, as the pruning loop requires a solution to \emph{many\/} such instances, one per $\smash{(\Consumer, \Bundle)}$ pair (Line \ref{alg:EAP:pruningloop} of \Cref{alg:EAP}).
Alternatively, one could compute looser upper bounds, and thereby trade off computation time for opportunities to prune more $\smash{(\Consumer, \Bundle)}$ pairs, when the upper bound is not tight enough.
In our experiments, we show that even relatively loose but cheap-to-compute upper bounds result in significant pruning and, thus, savings along both dimensions---computational and sample complexity.

To conclude this section, we establish the correctness of \EAP. For our proof we rely on the following generalization of the first welfare theorem of economics, which handles additive errors.

\begin{theorem}[First Welfare Theorem~\cite{roughgarden2010algorithmic}]
\label{thm:fwt}
For $\epsilon > 0$,
let $(\Allocation, \Pricing)$ be an $\epsilon$-competitive equilibrium of $\Market$. Then, $\Allocation$ is a welfare-maximizing allocation of $\Market$, up to additive error $\epsilon \NumberOfConsumers$. 
\end{theorem}

\begin{theorem}[Elicitation Algorithm  with Pruning Guarantees]
\label{thm:EAP}
Let $\ConditionalMarket{\SetOfConditions}$ be a conditional market, let $\ConditionDistribution$ be a distribution over $\SetOfConditions$, and let $\ValuationRange \in \mathbb{R}_+$.
Suppose that for all $\Condition \in \SetOfConditions$ and $(\Consumer, \Bundle) \in \IndexSet$, it holds that $\ValuationFunction{\Consumer} (\Bundle, \Condition) \in [0, \ValuationRange]$, where $\ValuationRange \in \mathbb{R}$.
Let $\SamplingSchedule$ be a sequence of strictly increasing integers, and $\DeltaSchedule$ a sequence of the same length as $\SamplingSchedule$ such that $\DeltaScheduleIndex{\PruningIterationIndex} \in (0, 1)$ and $\sum_{\PruningIterationIndex} \DeltaScheduleIndex{\PruningIterationIndex} \in (0, 1)$.
If \EAP{}
outputs \\
    $\smash{(\{\EmpiricalValuationFunction{\Consumer}\}_{\Consumer \in \SetOfConsumers}, 
    \{\hat{\epsilon}_{\Consumer, \Bundle}\}_{(\Consumer, \Bundle) \in \SetOfConsumers \times 2^\SetOfGoods},
    \smash{1 - \sum_\PruningIterationIndex \DeltaScheduleIndex{\PruningIterationIndex}},
    \hat{\epsilon})}$ 
on input 
    $\smash{(\ConditionalMarket{\SetOfConditions}, 
    \ConditionDistribution, 
    \SamplingSchedule, 
    \DeltaSchedule, 
    \ValuationRange,
    \epsilon)}$, then the following holds with probability at least $\smash{1 - \sum_\PruningIterationIndex \DeltaScheduleIndex{\PruningIterationIndex}}$:

    1. $\norm{\Market_{\ConditionDistribution} - \hat{\Market}}_{\IndexSet} \le \hat{\epsilon}_{\Consumer, \Bundle}$
    
    2. $\CESet(\ExpectedMarket{\ConditionDistribution}) 
        \subseteq 
            \CESet_{2 \hat{\epsilon}} (\hat{\Market}) 
            \subseteq 
            \CESet_{4 \hat{\epsilon}} (\ExpectedMarket{\ConditionDistribution})$%

Here $\hat{\Market}$ is the empirical market obtained via \EAP, i.e., the market with valuation functions given by $\{\EmpiricalValuationFunction{\Consumer}\}_{\Consumer \in \SetOfConsumers}$. 
\end{theorem}

\begin{proof}
To show part 1, note that at each iteration $\PruningIterationIndex$ of \EAP, Line~\ref{alg:EAP:errorUpdate} updates the error estimates for each $(\Consumer, \Bundle)$ after a call to \EA{} (Line~\ref{alg:EAP:callEA} of \EAP) with input failure probability $\DeltaScheduleIndex{\PruningIterationIndex}$. \Cref{thm:EAGuarantees} implies that each call to \EA{} returns estimated values that are within $\hat{\epsilon}$ of their expected value with probability at least $1 - \DeltaScheduleIndex{\PruningIterationIndex}$. By union bounding all calls to \EA{} within \EAP, part 1 then holds with probability at least $1 - \sum_\PruningIterationIndex \DeltaScheduleIndex{\PruningIterationIndex}$.

To show part 2, note that only pairs $(\Consumer, \Bundle)$ for which Equation~(\ref{eq:pruning}) holds are removed from index set $\IndexSet$ (Line~\ref{alg:EAP:pruning} of \EAP). By \Cref{lemma:pruning}, no such pair can be part of any approximate welfare-maximizing allocation of the expected market, $\smash{\ExpectedMarket{\ConditionDistribution}}$. By \Cref{thm:fwt}, no such pair can be a part of any CE. Consequently, $\smash{\hat{\Market}}$ contains accurate enough estimates (up to $\epsilon$) of all $(\Consumer, \Bundle)$ pairs that may participate in any CE. Part 2 then follows from \Cref{thm:CEApprox}.
\end{proof}

%% file: algos/eap.tex
\setlength{\textfloatsep}{5pt}%
\begin{algorithm}[tb]
\caption{Elicitation Algorithm with Pruning (\EAP)}
\label{alg:EAP}\vspace{0.35em}
\raggedright\textbf{Input}: 
    $\ConditionalMarket{\SetOfConditions}, 
    \ConditionDistribution, 
    \SamplingSchedule, 
    \DeltaSchedule, 
    \ValuationRange,
    \epsilon$.
    
A conditional combinatorial market $\ConditionalMarket{\SetOfConditions}$,
a distribution $\ConditionDistribution$ over $\SetOfConditions$,
a sampling schedule $\SamplingSchedule$,
a failure probability schedule $\DeltaSchedule$,
a pruning budget schedule $\PruningSchedule$,
a valuation range $\ValuationRange$, and
a target approx. error $\epsilon$. \\ \vspace{0.35em}
\textbf{Output}: 
Valuation estimates $\EmpiricalValuationFunction{\Consumer} (\Bundle)$, for all $(\Consumer, \Bundle)$,
approximation errors $\hat{\epsilon}_{\Consumer, \Bundle}$,
\mbox{failure probability $\hat{\delta}$, 
and CE error $\hat{\epsilon}$.}
\vspace{-1.em}
\begin{algorithmic}[1] %
\STATE $\IndexSet \gets \SetOfConsumers \times 2^\SetOfGoods$ 
    \quad\quad\quad \COMMENT{Initialize index set}
\STATE $(\EmpiricalValuationFunction{\Consumer} (\Bundle), \hat{\epsilon}_{\Consumer, \Bundle}) \gets
    (0, \nicefrac{c}{2}), \forall (\Consumer, \Bundle) \in \IndexSet$
    \COMMENT{Initialize outputs}
\FOR{$\PruningIterationIndex \in 1, \ldots, |\SamplingSchedule|$}
\STATE $(\{\EmpiricalValuationFunction{\Consumer}\}_{(\Consumer, \Bundle) \in \IndexSet}, \hat{\epsilon}) \gets \textsc{EA} (\ConditionalMarket{\SetOfConditions},
    \ConditionDistribution, 
    \IndexSet, 
    \NumberOfSamples_\PruningIterationIndex, 
    \delta_\PruningIterationIndex, 
    \ValuationRange)$ 
    \COMMENT{Call Alg. \ref{alg:EA}}
\label{alg:EAP:callEA}
\STATE $\hat{\epsilon}_{\Consumer, \Bundle} \gets 
            \hat{\epsilon}, \forall (\Consumer, \Bundle) \in \IndexSet$
    \quad\quad\quad \COMMENT{Update error rates}
\label{alg:EAP:errorUpdate}
\IF{$\hat{\epsilon} \le \epsilon$ or 
    $\PruningIterationIndex = |\SamplingSchedule|$ or 
    $\IndexSet = \emptyset$}
\STATE \textbf{return} 
    $(\{\EmpiricalValuationFunction{\Consumer}\}_{\Consumer \in \SetOfConsumers}, 
    \{\hat{\epsilon}_{\Consumer, \Bundle}\}_{(\Consumer, \Bundle) \in \SetOfConsumers \times 2^\SetOfGoods},
    \sum_{l=1}^\PruningIterationIndex \DeltaScheduleIndex{l},
    \hat{\epsilon})$
    \label{alg:EAP:termination}
\ENDIF
\STATE Let $\hat{\Market}$ be the market with valuations $\{\EmpiricalValuationFunction{\Consumer}\}_{(\Consumer, \Bundle) \in \IndexSet}$
\STATE $\IndexSet_{\textsc{prune}} \gets \emptyset$ 
    \quad\quad\quad \COMMENT{Initialize set of indices to prune}
\STATE $\IndexSet_{\textsc{candidates}} \gets $ a subset of $\IndexSet$ of size at most $\PruningScheduleIndex{\PruningIterationIndex}$

    $\quad\quad$\COMMENT{Select some active pairs as candidates for pruning}
\label{alg:EAP:candidates}
\FOR{$(\Consumer, \Bundle) \in \IndexSet_{\textsc{candidates}}$} \label{alg:EAP:pruningloop}
\STATE Let $\SubMarketHat$ be the $(\Consumer, \Bundle)$-submarket of $\hat{\Market}$.
\STATE Let $\WelfareUpBound{\Consumer}{\Bundle}$ an upper bound of $\Welfare^*(\SubMarketHat)$.
\label{alg:EAP:pruning}
\IF{$\smash{\EmpiricalValuationFunction{\Consumer} (\Bundle) +
        \WelfareUpBound{\Consumer}{\Bundle} + 2 \hat{\epsilon} \NumberOfConsumers
        < \Welfare^*(\hat{\Market})}$} \label{alg:EAP:pruningstep}
\STATE $\IndexSet_{\textsc{prune}} \gets \IndexSet_{\textsc{prune}} \cup (\Consumer, \Bundle)$
\ENDIF
\ENDFOR
\STATE $\IndexSet \gets \IndexSet \setminus \IndexSet_{\textsc{prune}}$
\ENDFOR
\end{algorithmic}
\end{algorithm}

%% file: algos/ea.tex
\setlength{\textfloatsep}{5pt}%
\begin{algorithm}[tb]
\caption{Elicitation Algorithm (\EA)}
\label{alg:EA} \vspace{0.35em}
\raggedright\textbf{Input}: 
    $\ConditionalMarket{\SetOfConditions}, 
    \ConditionDistribution, 
    \IndexSet, 
    \NumberOfSamples, 
    \delta, 
    \ValuationRange$.
    
A conditional combinatorial market $\ConditionalMarket{\SetOfConditions}$,
a distribution $\ConditionDistribution$ over $\SetOfConditions$,
an index set $\IndexSet$,
sample size $\NumberOfSamples$,
failure prob. $\delta$, and
valuation range $\ValuationRange$. \\ \vspace{0.35em}
\textbf{Output}:
Valuation estimates $\EmpiricalValuationFunction{\Consumer} (\Bundle)$, for all $(\Consumer, \Bundle) \in \IndexSet$, 
and an approximation error $\hat{\epsilon}$.
\begin{algorithmic}[1] %
\STATE $(\Condition_1, \ldots, \Condition_\NumberOfSamples) \sim \ConditionDistribution$ 
    \quad\quad\quad \COMMENT{Draw $\NumberOfSamples$ samples from $\ConditionDistribution$}
\FOR{$(\Consumer, \Bundle) \in \IndexSet$}
\STATE $\EmpiricalValuationFunction{\Consumer} (\Bundle) \gets \frac{1}{\NumberOfSamples} \sum_{\SampleIndex = 1}^\NumberOfSamples \ValuationFunction{\Consumer} (\Bundle, \Condition_\SampleIndex)$
\ENDFOR
\STATE $\hat{\epsilon} 
        \gets \ValuationRange
        \sqrt{\nicefrac{\ln(\nicefrac{2|\IndexSet|}{\delta})}{2\NumberOfSamples}}$
        \quad\quad\quad \COMMENT{Compute error}
\STATE \textbf{return} 
    $(\{\EmpiricalValuationFunction{\Consumer}\}_{(\Consumer, \Bundle) \in \IndexSet}, \hat{\epsilon})$
\end{algorithmic}
\end{algorithm}

%% file: equations/opt_welfare.tex
\begin{align*}
\Welfare^*(\Market)
= \sum_{\Consumer \in \SetOfConsumers}\ValuationFunction{\Consumer} (\Allocation^*_\Consumer) 
\ge 
\sum_{\Consumer \in \SetOfConsumers} \ValuationFunction{\Consumer} (\AllocationAlt^*_\Consumer)
\ge
\sum_{\Consumer \in \SetOfConsumers} \ValuationFunctionAlt{\Consumer} (\AllocationAlt^*_\Consumer) - \epsilon \NumberOfConsumers \\
= 
\Welfare^*(\MarketAlt) - \epsilon \NumberOfConsumers
\end{align*}

%% file: expts.tex
\section{Experiments}
\label{sec:expts}

The goal of our experiments is to robustly evaluate the empirical performance of our algorithms. To this end, we experiment with a variety of qualitatively different inputs. In particular, we evaluate our algorithms on both unit-demand valuations, the Global Synergy Value Model (GSVM)~\cite{goeree2010hierarchical}, and the Local Synergy Value Model (LSVM)~\cite{scheffel2012impact}. Unit-demand valuations are a class of valuations central to the literature on economics and computation~\cite{lehmann2006combinatorial} for which efficient algorithms exist to compute CE~\cite{gul1999walrasian}. GSVM and LSVM model situations in which buyers' valuations encode complements; CE are not known be efficiently computable, or even representable, in these markets.

While CE are always guaranteed to exist (e.g.,~\cite{bikhchandani2002package}),
in the worst case, they might require personalized bundle prices. These prices are computationally complex, not to mention out of favor~\cite{hinz2011price}.
A pricing $\Pricing = (\PriceFunction_1, \ldots, \PriceFunction_{\NumberOfConsumers})$ is \emph{anonymous} if it charges every buyer the same price, i.e., $\PriceFunction_\Consumer = \PriceFunction_\ConsumerAlt = \PriceFunction$ for all $\Consumer \ne \ConsumerAlt \in \SetOfConsumers$. Moreover, an anonymous pricing is \emph{linear} if there exists a set of prices $\{\Price_1, \ldots, \Price_\NumberOfGoods\}$, where $\Price_\Good$ is good $\Good$'s price, such that $\PriceFunction(\Bundle) = \sum_{\Good \in \Bundle} \Price_\Good$. In what follows, we refer to linear, anonymous pricings as linear prices. 

Where possible, it is preferable to work with linear prices, as they are simpler, e.g., when bidding in an auction~\cite{kwasnica2005new}.
In our present study---one of the first
empirical studies on learning CE---we thus focus on linear prices, leaving as future research the empirical%
\footnote{Note that all our theoretical results hold for any pricing profile.} effect of more complex pricings.%
\footnote{\citeauthor{lahaie2019adaptive}~\cite{lahaie2019adaptive}, for example, search for prices in between linear and bundle.}

To our knowledge, there have been no analogous attempts at learning CE;
hence, we do not reference any baseline algorithms from the literature.
Rather, we compare the performance of \EAP, our pruning algorithm, to \EA, investigating the quality of the CE learned by both, as well as their sample efficiencies.

\subsection{Experimental Setup.}
We first explain our experimental setup, and then present results.
We let $U[a, b]$ denote the continuous uniform distribution over range $[a, b]$, and $U\{k, l\}$, the discrete uniform distribution over set $\{ k, k+1, \ldots, l \}$, for $k \le l \in \mathbb{N}$.

\paragraph{Simulation of Noisy Combinatorial Markets.}
We start by drawing markets from experimental market
distributions. Then, fixing a market, we simulate noisy value elicitation by adding noise drawn from experimental noise distributions to buyers' valuations in the market.
We refer to a market realization $\Market$ drawn from an experimental market distribution as the \emph{ground-truth} market. Our experiments then measure how well we can approximate the CE of a ground-truth market $\Market$ given access only to noisy samples of it.

Fix a market $\Market$ and a condition set $\SetOfConditions = [a, b]$, where $a < b$.
Define the conditional market $\ConditionalMarket{\SetOfConditions}$, where $\Valuation_{\Consumer}(\Bundle, \Condition_{\Consumer \Bundle}) =  \Valuation_{\Consumer}(\Bundle) +  \Condition_{\Consumer \Bundle}$, for $\Condition_{\Consumer \Bundle} \in \SetOfConditions$. In words, when eliciting $\Consumer$'s valuation for $\Bundle$, we assume additive noise, namely $\Condition_{\Consumer \Bundle}$. The market $\ConditionalMarket{\SetOfConditions}$ together with distribution $\ConditionDistribution$ over $\SetOfConditions$ is the model from which our algorithms elicit noisy valuations from buyers.  
Then, given samples $\VectorOfConditions$ of $\ConditionalMarket{\SetOfConditions}$, the empirical market $\smash{\EmpiricalMarket{\VectorOfConditions}}$ is the market estimated from the samples. Note that $\smash{\EmpiricalMarket{\VectorOfConditions}}$ is the only market we get to observe in practice. 

We consider only zero-centered noise distributions. In this case, the expected combinatorial market $\ExpectedMarket{\ConditionDistribution}$ is the same as the ground-truth market $\Market$ since, for every $\Buyer, \Bundle \in \SetOfConsumers \times 2^\SetOfGoods$ it holds that $\ExpectedValuationFunction{\Consumer} (\Bundle, \ConditionDistribution) = \Expectation_{\ConditionDistribution}[\Valuation_{\Consumer}(\Bundle, \Condition_{\Consumer \Bundle})] = \Expectation_{\ConditionDistribution}[\Valuation_{\Consumer}(\Bundle) +  \Condition_{\Consumer \Bundle}] = \Valuation_{\Consumer}(\Bundle)$.
While this noise structure is admittedly simple, we robustly evaluate our algorithms along another dimension, as we study several rich market structures (unit-demand, GSVM, and LSVM). An interesting future direction would be to also study richer noise structures, e.g., letting noise vary with a bundle's size, or other market characteristics.

\paragraph{Utility-Maximization (\UM) Loss}
To measure the quality of a CE $(\alternative{\Allocation}, \alternative{\Pricing})$ computed for a market $\alternative{\Market}$ in another market $\Market$, we first define the per-buyer metric $\UMLoss_{\Market, \Consumer}$ as follows, $$\UMLoss_{\Market, \Consumer} (\alternative{\Allocation}, \alternative{\Pricing})= \smash{\max_{\Bundle \subseteq \SetOfGoods} ( \Valuation_\Consumer (\Bundle) - \alternative{\PriceFunction}(\Bundle) ) - (\Valuation_\Consumer (\alternative{\Bundle_\Consumer}) - \alternative{\PriceFunction} (\alternative{ \Bundle_\Consumer}))},$$ i.e., the difference between the maximum utility $\Consumer$ could have attained at prices $\alternative{\Pricing}$ and the utility $\Consumer$ attains at the outcome $(\alternative{\Allocation}, \alternative{\Pricing})$. Our metric of interest is then $\UMLoss_\Market$ defined as, $$\UMLoss_\Market (\alternative{\Allocation}, \alternative{\Pricing}) = \max_{\Consumer \in \SetOfConsumers} \UMLoss_{\Market, \Consumer}(\alternative{\Allocation}, \alternative{\Pricing}),$$ which is a worst-case measure of utility loss over all buyers in the market. Note that it is not useful to incorporate the SR condition into a loss metric, because it is always satisfied.

In our experiments, we measure the \UM{} loss that a CE of an empirical market
obtains, evaluated in the corresponding ground-truth market. Thus, given an empirical estimate $\smash{\EmpiricalMarket{\VectorOfConditions}}$ of $\Market$, and a CE $(\hat \Allocation, \hat \Pricing)$ in $\smash{\EmpiricalMarket{\VectorOfConditions}}$, we measure $\smash{\UMLoss_\Market(\hat \Allocation, \hat \Pricing)}$, i.e., the loss in $\Market$ at prices $\hat \Pricing$ of CE $\smash{(\hat \Allocation, \hat \Pricing)}$.
\Cref{thm:CEApprox} implies that if $\smash{\EmpiricalMarket{\VectorOfConditions}}$ is an $\epsilon$-approximation of $\Market$, then $\smash{\UMLoss_\Market(\hat \Allocation, \hat \Pricing) \le 2 \epsilon}$. Moreover, \Cref{thm:EAGuarantees} yields the same guarantees, but with probability at least $1 - \delta$, provided the $\epsilon$-approximation holds with probability at least $1 - \delta$.

\paragraph{Sample Efficiency of \EAP.}
We say that algorithm $A$ has better \mydef{sample efficiency} than algorithm $B$ if $A$ requires fewer samples than $B$ to achieve at least the same $\epsilon$ accuracy.

Fixing a condition set $\SetOfConditions$, a distribution $\ConditionDistribution$ over $\SetOfConditions$, and a conditional market $\ConditionalMarket{\SetOfConditions}$, we use the following experimental design to evaluate \EAP's sample efficiency relative to that of \EA. 
Given a desired error guarantee $\epsilon > 0$, we compute the number of samples $\NumberOfSamples(\epsilon)$ that would be required for \EA{} to achieve accuracy $\epsilon$. We then use the following doubling strategy as a sampling schedule for \EAP, $\SamplingSchedule(\NumberOfSamples(\epsilon)) = [ \nicefrac{\NumberOfSamples(\epsilon)}{4}, \nicefrac{\NumberOfSamples(\epsilon)}{2}, \NumberOfSamples(\epsilon), 2\NumberOfSamples(\epsilon)]$, rounding to the nearest integer as necessary,
and the following failure probability schedule $\DeltaSchedule = [0.025, 0.025, 0.025, 0.025]$,
which sums to $0.1$.

Finally, the exact pruning budget schedules will vary depending on the value model (unit demand, GSVM, or LSVM). But in all cases, we denote an \emph{unconstrained} pruning budget schedule by $\PruningSchedule = [\infty, \infty, \infty, \infty]$, which by convention means that at every iteration, all active pairs are candidates for pruning.
Using these schedules, we run \EAP{} with a desired accuracy of zero. We denote by $\epsilon_{\EAP} (\epsilon)$ the approximation guarantee achieved by \EAP{} upon termination.

\subsection{Unit-demand Experiments}
A buyer $\Consumer$ is endowed with unit-demand valuations if, for all $\Bundle \subseteq \SetOfGoods$, $\ValuationFunction{\Consumer}(\Bundle) = \max_{\Good \in \Bundle} \Valuation_{\Consumer} (\{\Good\})$.
In a unit-demand market, all buyers have unit-demand valuations.
A unit-demand market can be compactly represented by matrix $\UnitDemandMatrix$, where entry $\Valuation_{\Consumer \Good} \in \mathbb{R}_+$ is $\Consumer$'s value for $\Good$, i.e., $\Valuation_{\Consumer \Good} = \Valuation_{\Consumer} (\{\Good\})$.
In what follows, we denote by $\UnitDemandMatrixRV$ a random variable over unit-demand valuations.

We construct four different distributions over unit-demand markets: \textsc{Uniform}, \textsc{Preferred-Good}, \textsc{Preferred-Good-Distinct}, and \textsc{Preferred-Subset}.
All distributions are parameterized by $\NumberOfConsumers$ and $\NumberOfGoods$, the number of buyers and goods, respectively. 
A uniform unit-demand market $\UnitDemandMatrixRV \sim \textsc{Uniform}$ is such that for all $\Consumer, \Good,$ $\Valuation_{\Consumer \Good} \sim U[0, 10]$. 
When $\UnitDemandMatrixRV \sim \textsc{Preferred-Good}$, each buyer $\Consumer$ has a preferred good $\Good_\Consumer$, with $\Good_\Consumer \sim U\{1, \ldots, \NumberOfGoods\}$ and  $\Valuation_{\Consumer \Good_\Consumer} \sim U[0, 10]$.
Conditioned on $\Valuation_{\Consumer \Good_\Consumer}$, $\Consumer$'s value for good $k \neq \Good_\Consumer$ is given by $\Valuation_{\Consumer k} = \nicefrac{\Valuation_{\Consumer \Good_\Consumer}}{2^k}$.
Distribution $\textsc{Preferred-Good-Distinct}$ is similar to $\textsc{Preferred-Good}$, except that no two buyers have the same preferred good.
(Note that the $\textsc{Preferred-Good-Distinct}$ distribution is only well defined when $\NumberOfConsumers \leq \NumberOfGoods$.)
Finally, 
when $\UnitDemandMatrixRV \sim \textsc{Preferred-Subset}$, each buyer $\Consumer$ is interested in a subset of goods $\Consumer_\SetOfGoods \subseteq \SetOfGoods$, where $\Consumer_\SetOfGoods$ is drawn uniformly at random from the set of all bundles.
Then, the value $\Consumer$ has for $\Good$ is given by $\Valuation_{\Consumer \Good} \sim U[0, 10]$, if $\Good \in \Consumer_\SetOfGoods$; and 0, otherwise.

In unit-demand markets, we experiment with three noise models, low, medium, and high, by adding noise drawn from $U[-.5, .5]$, $U[-1, 1],$ and $U[-2, 2]$, respectively.
We choose $\NumberOfConsumers, \NumberOfGoods \in \{5, 10, 15, 20\}^2$.

\input{tables/um_loss}
\input{plots/unit_demand_heatmap}

\paragraph{Unit-demand Empirical \UM{} Loss of \EA.}

As a learned CE is a CE of a learned market, we require a means of computing the CE of a market---specifically, a unit-demand market $\UnitDemandMatrix$. To do so, we first solve for the%
\footnote{Since in all our experiments, we draw values from continuous distributions, we assume that the set of markets with multiple welfare-maximizing allocations is of negligible size. Therefore, we can ignore ties.} welfare-maximizing allocation $\Allocation^*_\UnitDemandMatrix$ of $\UnitDemandMatrix$, by solving for the maximum weight matching using Hungarian algorithm~\cite{kuhn1955hungarian} in the bipartite graph whose weight matrix is given by $\UnitDemandMatrix$. Fixing $\Allocation^*_\UnitDemandMatrix$, we then solve for prices via linear programming~\cite{bikhchandani2002package}.
In general, there might be many prices that couple with $\Allocation^*_\UnitDemandMatrix$ to form a CE of $\UnitDemandMatrix$. For simplicity, we solve for two pricings given $\Allocation^*_\UnitDemandMatrix$, the revenue-maximizing $\Pricing_{\textsc{max}}$ and revenue-minimizing $\Pricing_{\textsc{min}}$, where revenue is defined as the sum of the prices.

For each distribution, we draw 50 markets, and for each such market $\UnitDemandMatrix$, we run \EA{} four times, each time to achieve guarantee $\epsilon \in \{0.05, 0.1, 0.15, 0.2\}$. \EA{} then outputs an empirical estimate $\hat{\UnitDemandMatrix}$ for each $\UnitDemandMatrix$. We compute outcomes $\smash{(\Allocation^*_{\hat{\UnitDemandMatrix}}, \hat{\Pricing}_{\textsc{max}})}$ and $\smash{(\Allocation^*_{\hat{\UnitDemandMatrix}}, \hat{\Pricing}_{\textsc{min}})}$, and measure $\smash{\UMLoss_\UnitDemandMatrix(\Allocation^*_{\hat{\UnitDemandMatrix}}, \hat{\Pricing}_{\textsc{max}})}$ and $\smash{\UMLoss_\UnitDemandMatrix(\Allocation^*_{\hat{\UnitDemandMatrix}}, \hat{\Pricing}_{\textsc{min}})}$.
We then average across all market draws, for both the minimum and the maximum pricings.
\Cref{tab:EAResults} summarizes a subset of these results. %
The error guarantees are consistently met across the board, indeed by one or two orders of magnitude, and they degrade as expected: i.e., with higher values of $\epsilon$.
We note that the quality of the learned CE is roughly the same for all distributions, except in the case of $\hat{\Pricing}_{\textsc{min}}$ and \textsc{Preferred-Good-Distinct}, where learning is more accurate. For this distribution, it is enough to learn the preferred good of each buyer. Then, one possible CE is to allocate each buyer its preferred good and price all goods at zero
which yields near no \UMLoss. Note that, in general, pricing all goods at zero is not a CE, unless the market has some special structure, like the markets drawn from \textsc{Preferred-Good-Distinct}. 

\vspace{-0.5em}
\paragraph{Unit-demand Sample Efficiency}
We use pruning schedule $\PruningSchedule = [\infty, \infty, \infty, \infty]$ and for each $(\Consumer, \Good)$ pair, we use the Hungarian algorithm~\cite{kuhn1955hungarian} to compute the optimal welfare of the market without $(\Consumer, \Good)$. In other words, in each iteration, we consider all active $(\Consumer, \Good)$ pairs as pruning candidates (\Cref{alg:EAP}, Line \ref{alg:EAP:candidates}), and for each we compute the optimal welfare (\Cref{alg:EAP}, Line \ref{alg:EAP:pruning}).

For each market distribution, we compute the average of the number of samples used by \EAP{} across 50 independent market draws. We report samples used by \EAP{} as a percentage of the number of samples used by \EA{} to achieve the same guarantee, namely, $\epsilon_{\EAP} (\epsilon)$, for each initial value of $\epsilon$.
\Cref{fig:EAPSampleEff_unit_demand} depicts the results of these experiments as heat maps, for all distributions and for $\epsilon = 0.05$, where darker colors indicate more savings, and thus better \EAP{} sample efficiency.

A few trends arise, which we note are similar for other values of $\epsilon$.
For a fixed number of buyers, \EAP's sample efficiency improves as the number of goods increases, because fewer goods can be allocated, which means that there are more candidate values to prune, resulting in more savings.
On the other hand,
the sample efficiency usually decreases as the number of buyers increases; this is to be expected, as the pruning criterion degrades with the number of buyers (\Cref{lemma:pruning}).
While savings exceed 30\% across the board, we note that \textsc{Uniform}, the market with the least structure, achieves the least savings, while \textsc{Preferred-Subset} and \textsc{Preferred-Good-Distinct} achieve the most.
This finding shows that \EAP{} is capable of exploiting the structure present in these distributions, despite not knowing anything about them \emph{a priori}.

Finally, we note that sample efficiency quickly degrades for higher values of $\epsilon$.
In fact, for high enough values of $\epsilon$ (in our experiments, $\epsilon = 0.2$), \EAP{} might, on average, require more samples than \EA{} to produce the same guarantee.
Most of the savings achieved are the result of pruning enough $(\Consumer, \Good)$ pairs early enough: i.e., during the first few iterations of \EAP.
When $\epsilon$ is large, however,
our sampling schedule does not allocate enough samples early on.
When designing sampling schedules for \EAP, one must allocate enough (but not too many) samples at the beginning of the schedule. Precisely how to determine this schedule is an empirical question, likely dependent on the particular application at hand.

%% file: tables/um_loss.tex
\begin{table}
\noindent\resizebox{\columnwidth}{!}{
\begin{tabular}{l|rr|rr}
\toprule
&\multicolumn{2}{|c}{$\epsilon = 0.05$} &\multicolumn{2}{|c}{$\epsilon = 0.2$} \\
Distribution & 
$\hat{\Pricing}_{\textsc{min}}$ & $\hat{\Pricing}_{\textsc{max}}$ & $\hat{\Pricing}_{\textsc{min}}$ & $\hat{\Pricing}_{\textsc{max}}$\\
\midrule
                 \textsc{Uniform} &  0.0018 &  0.0020 &  0.0074 &  0.0082 \\
          \textsc{Preferred-Good} &  0.0019 &  0.0023 &  0.0080 &  0.0094 \\
 \textsc{Preferred-Good-Distinct} &  0.0000 &  0.0020 &  0.0000 &  0.0086 \\
        \textsc{Preferred-Subset} &  0.0019 &  0.0022 &  0.0076 &  0.0090 \\
\bottomrule
\end{tabular}
}
\caption{Average \UMLoss{} for $\epsilon \in \{0.05, 0.2\}$.}
\label{tab:EAResults}
\vspace{-5mm}
\end{table}

%% file: plots/unit_demand_heatmap.tex
\begin{figure*}[!ht]
\includegraphics[width=1.0\textwidth]{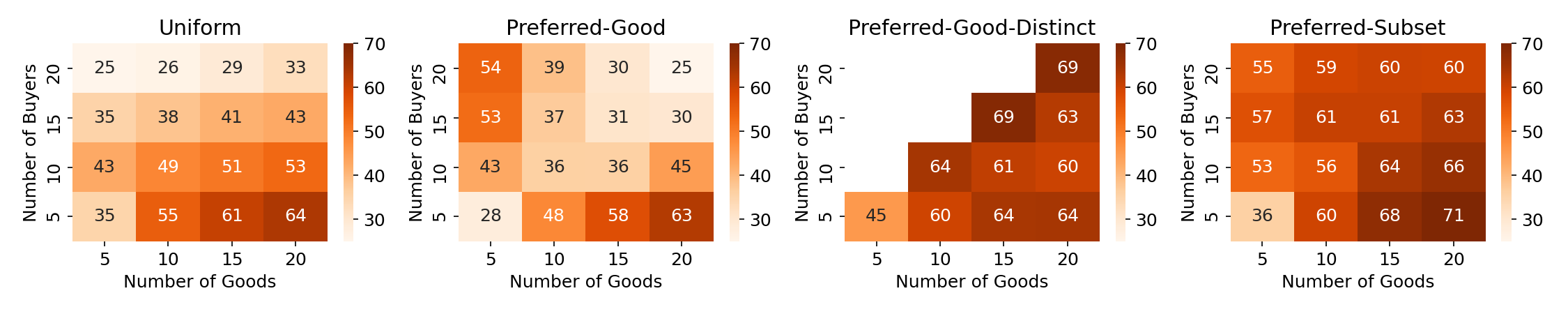}
    \vspace{-8mm}
    \caption{Mean \EAP{} sample efficiency relative to \EA{}, $\epsilon = 0.05$. 
    Each $(\Consumer, \Good)$ pair is annotated with the corresponding \% saving. }
  \label{fig:EAPSampleEff_unit_demand}
\vspace{-2mm}
\end{figure*}

%% file: value_models.tex
\vspace{-.5em}
\subsection{Value Models}

In this next set of experiments, we test the empirical performance of our algorithms in 
more complex markets, where buyers valuations contain synergies. Synergies are a common feature of many high-stakes combinatorial markets. For example, telecommunication service providers might value different bundles of radio spectrum licenses differently, depending on whether the licenses in the bundle complement one another. For example, a bundle including New Jersey and Connecticut might not be very valuable unless it also contains New York City. 

Specifically, we study the Global Synergy Value Model (GSVM)~\cite{goeree2010hierarchical} and the Local Synergy Value Model (LSVM)~\cite{scheffel2012impact}. These models or markets capture buyers' synergies as a function of buyers' types and their (abstract) geographical locations. In both GSVM and LSVM, there are 18 licenses, with buyers of two types: national or regional. A national buyer is interested in larger packages than regional buyers, whose interests are limited to certain regions. GSVM has six regional bidders and one national bidder and models geographical regions as two circles. LSVM has five regional bidders and one national bidder and uses a rectangular model. The models differ in the exact ways buyers' values are drawn, but in any case, synergies are modeled by suitable distance metrics.
In our experiments, we draw instances of both GSVM and LSVM using SATS, a universal spectrum auction test suite developed by researchers to test algorithms for combinatorial markets~\cite{weiss2017sats}.

\input{tables/value_models_vertical}

\vspace{-0.5em}
\paragraph{Experimental Setup.}
On average, the value a buyer has for an arbitrary bundle in either GSVM or LSVM markets is approximately 80. We introduce noise i.i.d. noise from distribution $U[-1, 1]$ whose range is 2, or 2.5\% of the expected buyer's value for a bundle.
As GSVM's buyers' values are at most 400, and LSVM's are at most 500, we use valuation ranges $\ValuationRange = 402$ and $\ValuationRange = 502$ for GSVM and LSVM, respectively.
We note that a larger noise range yields qualitatively similar results with errors scaling accordingly.

For the GSVM markets, we use the pruning budget schedule $\PruningSchedule = [\infty, \infty, \infty, \infty]$.
For each $(\Consumer, \Bundle)$ pair, we solve the welfare maximization problem using an off-the-shelf solver.%
\footnote{We include ILP formulations and further technical details in the appendix.}
In an LSVM market, the national bidder demands all 18 licenses. The welfare optimization problem in an LSVM market is solvable in a few seconds.%
\footnote{Approximately 20 seconds in our experiments, details appear in the appendix.} Still, the many submarkets (in the hundreds of thousands) call for a finite pruning budget schedule and a cheaper-to-compute welfare upper bound.
In fact, to address LSVM's size complexity, we slightly modify \EAP{}, as explained next.

\paragraph{A two-pass strategy for LSVM}
Because of the complexity of LSVM markets, we developed a heuristic pruning strategy, in which we perform two pruning passes during each iteration of \EAP. The idea is to compute a computationally cheap upper bound on welfare with pruning budget schedule $\PruningSchedule = [\infty, \infty, \infty, \infty]$ in the first pass, use this bound instead of the optimal for each active $(\Consumer, \Bundle)$. We compute this bound using the classic relaxation technique to create admissible heuristics.
Concretely, given a candidate $(\Consumer, \Bundle)$ pair, we compute the maximum welfare in the absence of pair $(\Consumer, \Bundle)$, ignoring feasibility constraints:
\vspace{-0.1em}
$$\WelfareUpBound{\Consumer}{\Bundle} = \sum\nolimits_{\ConsumerAlt \in \SetOfConsumers \setminus \{\Consumer\}} 
    \max\{
        \Valuation_{\ConsumerAlt}(\BundleAlt) 
            \mid 
        \BundleAlt \in 2^\SetOfGoods \text{ and } 
        \Bundle \cap \BundleAlt = \emptyset\}
\vspace{-0.1em}
$$

After this first pass, we undertake a second pass over all remaining active pairs. For each active pair, we compute the optimal welfare without pair $(\Consumer, \Bundle)$, but using the following \emph{finite\/} pruning budget schedule $\PruningSchedule =  [180, 90, 60, 45]$. In other words, we carry out this computation for just a few of the remaining candidate pairs. We chose this pruning budget schedule so that one iteration of \EAP{} would take approximately two hours.

One choice remains undefined for the second pruning pass: which $(\Consumer, \Bundle)$ candidate pairs to select out of those not pruned in the first pass? %
For each iteration $\PruningIterationIndex$, we sort the $(\Consumer, \Bundle)$ in descending order according to the upper bound on welfare computed in the first pass, and then we select the bottom $\PruningSchedule_\PruningIterationIndex$ pairs (180 during the first iteration, 90 during the second, etc.). The intuition for this choice is that pairs with lower upper bounds might be more likely to satisfy \Cref{lemma:pruning}'s pruning criteria than pairs with higher upper bounds. Note that the way candidate pairs are selected for the second pruning pass uses no information about the underlying market, and is thus widely applicable.
We will have more to say about the lack {\it a priori} information used by \EAP{} in what follows.

\paragraph{Results.}
\Cref{tab:value_models_results_ver} summarizes the results of our experiments with GSVM and LSVM markets. The table shows 95\% confidence intervals around the mean number of samples needed by \EA{} and \EAP{} to achieve the indicated accuracy ($\epsilon$) guarantee for each row of the table. The table also shows confidence intervals around the mean $\epsilon$ guarantees achieved by \EAP, denoted $\epsilon_{\EAP}$, and confidence intervals over the \UM{} loss metric. Several observations follow. 

Although ultimately a heuristic method, on average \EAP{} uses far fewer samples than \EA{} and produces significantly better $\epsilon$ guarantees. We emphasize that \EAP{} is capable of producing these results without any {\it a priori\/} knowledge about the underlying market. Instead, \EAP{} \emph{autonomously\/}
samples those quantities that can provably be part of an optimal solution.
The \EAP{} guarantees are slightly worse in the LSVM market than for GSVM, where we prune \emph{all\/} eligible $(\Consumer, \Bundle)$ pairs. 
In general, there is a tradeoff between computational and sample efficiency: at the cost of more computation, to find more pairs to prune up front, one can save on future samples.
Still, even with a rather restricted pruning budget $\PruningSchedule =  [180, 90, 60, 45]$ (compared to hundreds of thousands potentially active $(\Consumer, \Bundle)$ pairs), \EAP{} achieves substantial savings compared to \EA{} in the LSVM market. 

Finally, the \UM{} loss metric follows a trend similar to those observed for unit-demand markets, i.e., the error guarantees are consistently met and degrade as expected (worst guarantees for higher values of $\epsilon$). Note that in our experiments, all 40 GSVM market instances have equilibria with linear and anonymous prices.
In contrast, only 18 out of 32 LSVM market do, so the table reports \UM{} loss over this set. For the remaining 32 markets, we report here a \UM{} loss of approximately $12 \pm 4$ \emph{regardless\/} of the value of $\epsilon$.  This high \UM{} loss is due to the lack of CE in linear pricings which dominates any \UM{} loss attributable to the estimation of values.

%% file: tables/value_models_vertical.tex
\begin{table*}
\centering
\vspace{-1.5em}
\resizebox{!}{1.5cm}{
\begin{tabular}{c|rrrr|rrrr}
\toprule
    & \multicolumn{4}{|c}{GSVM} 
    & \multicolumn{4}{|c}{LSVM} \\
    $\epsilon$
    & \EA{} & \EAP{} & $\epsilon_{\EAP}$ & UM Loss
    & \EA{} & \EAP{} & $\epsilon_{\EAP}$ & UM Loss\\    
\midrule

1.25    & $2,642$ & ${\bf 720 \pm 10}$ &$0.73 \pm 0.01$ & $0.0022 \pm 0.0002$ 
        & $330,497 \pm 386$ & ${\bf 270,754 \pm 14,154}$ &$0.89 \pm 0.00$ & $0.0011 \pm 0.0003$ 
        \\
2.50    & $660$ & ${\bf 226 \pm 10}$ &$1.57 \pm 0.02$ & $0.0041 \pm 0.0005$ 
        & $82,624 \pm 96$ & ${\bf 73,733 \pm 3,629}$ &$1.78 \pm 0.00$ & $0.0018 \pm 0.0003$
        \\
5.00    & $165$ & ${\bf 117 \pm 11}$ &$3.41 \pm 0.03$ & $0.0063 \pm 0.0008$ 
        & ${\bf 20,656 \pm 24}$ & $22,054 \pm 933$ &$3.59 \pm 0.01$ & $0.0037 \pm 0.0005$ 
        \\
10.0    & ${\bf 41}$ & $69 \pm 4$ &$7.36 \pm 0.04$ & $0.0107 \pm 0.0010$ 
        & ${\bf 5,164 \pm 6}$ & $7,580 \pm 211$ &$7.27 \pm 0.01$ & $0.0072 \pm 0.0011$ 
        \\
\bottomrule
\end{tabular}
}
\caption{
GSVM (left group) and LSVM (right group) results. Each group reports sample efficiency and UM loss. Each row of the table reports results for a fixed value of $\epsilon$. Results are 95\% confidence intervals over 40 GSVM market draws and 50 LSVM market draws, except for \EA's number of samples in the case of GSVM which is a deterministic quantity (a GSVM market is of size 4,480). The values in {\bf bold} indicate the more sample efficient algorithm. Numbers of samples are reported in millions. 
\vspace{-0.25em}
}
\vspace{-2em}
\label{tab:value_models_results_ver}
\end{table*}

%% file: conc.tex
\section{Conclusion and Future Directions}
\label{sec:conc}

In this paper, we define noisy combinatorial markets as a model of combinatorial markets in which buyers'
valuations are not known with complete certainty, but noisy samples can be obtained, for example, by using approximate methods, heuristics, or truncating the run-time of a complete algorithm. 
For this model, we tackle the problem of learning CE.
We first show tight lower- and upper-bounds on the buyers' utility loss, and hence the set of CE, given a uniform approximation of one market by another.
We then develop learning algorithms that, with high probability, learn said uniform approximations using only finitely many samples.
Leveraging the first welfare theorem of economics, we define a pruning criterion under which an algorithm can provably stop learning about buyers' valuations for bundles, without affecting the quality of the set of learned CE.
We embed these conditions in an algorithm that we show experimentally is capable of learning CE with far fewer samples than a baseline.
Crucially, the algorithm need not know anything about this structure \mydef{a priori}; our algorithm is general enough to work in any combinatorial market.
Moreover, we expect substantial improvement with sharper sample complexity bounds; in particular, variance-sensitive bounds can be vastly more efficient when the variance is small, whereas Hoeffding's inequality essentially assumes the worst-case variance.

%% file: ack.tex
\paragraph{Acknowledgements}
This work was supported by NSF Award CMMI-1761546 and by DARPA grant FA8750.

%% file: appendix.tex
\vspace*{-2.25cm}    
\section*{Appendix}

\subsection*{Theoretical Proofs}

\begin{proof}(\Cref{lemma:hoeffding} of the main paper)

Let $\ConditionalMarket{\SetOfConditions}$ be a conditional combinatorial market, $\ConditionDistribution$ a distribution over $\SetOfConditions$, and $\IndexSet \subseteq \SetOfConsumers \times 2^\SetOfGoods$ an index set. 
Let $\VectorOfConditions = (\Condition_1, \ldots, \Condition_\NumberOfSamples) \sim \ConditionDistribution$ be a vector of $\NumberOfSamples$ samples drawn from $\ConditionDistribution$. 
Suppose that for all $\Condition \in \SetOfConditions$ and $(\Consumer, \Bundle) \in \IndexSet$, it holds that $\ValuationFunction{\Consumer} (\Bundle, \Condition) \in [0, \ValuationRange]$ where $\ValuationRange \in \mathbb{R}_+$. 
Let $\delta > 0$ and $\epsilon > 0$. 
Then, by Hoeffding's inequality~\cite{hoeffding1994probability},

\begin{equation}
\label{eq:hoeff}
Pr(
    |\Valuation_\Consumer(\Bundle) 
        - 
    \EmpiricalValuationFunction{\Consumer} (\Bundle)|
        \ge 
    \epsilon
    )
    \le  2e^{-2 \NumberOfSamples (\frac{\epsilon}{\ValuationRange})^2}
\end{equation}

Now, applying union bound over all events 
$|\Valuation_\Consumer(\Bundle) 
        - 
    \EmpiricalValuationFunction{\Consumer} (\Bundle)|
        \ge 
    \epsilon$ 
where $(\Consumer, \Bundle) \in \IndexSet$,

\begin{equation}
\label{eq:unionbound}
Pr\left(\bigcup_{(\Consumer, \Bundle) \in \IndexSet}
|\Valuation_\Consumer(\Bundle) 
        - 
    \EmpiricalValuationFunction{\Consumer} (\Bundle)|
        \ge 
    \epsilon
\right) 
\le 
\sum_{(\Consumer, \Bundle) \in \IndexSet}
Pr\left(
 |\Valuation_\Consumer(\Bundle) 
        - 
    \EmpiricalValuationFunction{\Consumer} (\Bundle)|
        \ge 
    \epsilon
\right)
\end{equation}

Using bound (\ref{eq:hoeff}) in the right-hand side of (\ref{eq:unionbound}), 

\begin{equation}
\label{eq:hoeffplusunionbound}
Pr\left(\bigcup_{(\Consumer, \Bundle) \in \IndexSet}
|\Valuation_\Consumer(\Bundle) 
        - 
    \EmpiricalValuationFunction{\Consumer} (\Bundle)|
        \ge 
    \epsilon
\right) 
\le 
\sum_{(\Consumer, \Bundle) \in \IndexSet}
    2e^{-2 \NumberOfSamples (\frac{\epsilon}{\ValuationRange})^2}
    =2|\IndexSet|e^{-2 \NumberOfSamples (\frac{\epsilon}{\ValuationRange})^2}
\end{equation}

Where the last equality follows because the summands on the right-hand size of \cref{eq:hoeffplusunionbound} do not depend on the summation index.
Now, note that \cref{eq:hoeffplusunionbound} implies a lower bound for the event that complements $\bigcup_{(\Consumer, \Bundle) \in \IndexSet}
|\Valuation_\Consumer(\Bundle) 
        - 
    \EmpiricalValuationFunction{\Consumer} (\Bundle)|
        \ge 
    \epsilon$,

\begin{equation}
\label{eq:hoeffplusunionboundfinal}
Pr\left(\bigcap_{(\Consumer, \Bundle) \in \IndexSet}
|\Valuation_\Consumer(\Bundle) 
        - 
    \EmpiricalValuationFunction{\Consumer} (\Bundle)|
        \le 
    \epsilon
\right) 
\ge 
    1 - 2|\IndexSet|e^{-2 \NumberOfSamples (\frac{\epsilon}{\ValuationRange})^2}
\end{equation}

The event 
$\bigcap_{(\Consumer, \Bundle) \in \IndexSet}
|\Valuation_\Consumer(\Bundle) 
        - 
    \EmpiricalValuationFunction{\Consumer} (\Bundle)|
        \le 
    \epsilon$
is equivalent to the event 
$\max_{(\Consumer, \Bundle) \in \IndexSet} 
            | \ValuationFunction{\Consumer} (\Bundle) - \EmpiricalValuationFunction{\Consumer} (\Bundle)| \le \epsilon$.
Setting $\delta = 2|\IndexSet|e^{-2 \NumberOfSamples (\frac{\epsilon}{\ValuationRange})^2}$ and solving for $\epsilon$ yields $\epsilon = \ValuationRange \sqrt{\nicefrac{\ln\left(\nicefrac{2|\IndexSet|}{\delta}\right)}{2 \NumberOfSamples}}$. 

The results follows by substituting $\epsilon$ in \cref{eq:hoeffplusunionboundfinal}.

\end{proof}

\subsection*{Mathematical Programs}
For our experiments, we solve for CE in linear prices. To compute CE in linear prices, we first solve for a welfare-maximizing allocation $\Allocation^*$ and then, fixing $\Allocation^*$, we solve for CE linear prices. 
Note that, if a CE in linear prices exists, then it is supported by any welfare-maximizing allocation~\cite{roughgarden2015prices}. Moreover, since valuations in our experiments are drawn from continuous distributions, we assume that the set of welfare-maximizing allocations for a given market is of negligible size. 

Next, we present the mathematical programs we used to compute welfare-maximizing allocations and find linear prices. Given a combinatorial market $\Market$, the following integer linear program, (\ref{welfare-max-ILP}), computes a welfare-maximizing allocation $\Allocation^*$. 
Note this formulation is standard in the literature~\cite{nisanalgorithmic}. 

\begin{equation}
\label{welfare-max-ILP}
\def\arraystretch{0.5}
\begin{array}{lrlll}
\text{maximize}  & 
    \displaystyle\sum\limits_{\Consumer \in \SetOfConsumers, \Bundle\subseteq\SetOfGoods}  \Valuation_{\Consumer}(\Bundle)x_{\Consumer\Bundle} \\\\
\text{subject to}
    
    & \displaystyle\sum\limits_{\Consumer \in \SetOfConsumers, \Bundle \mid \Good \in \Bundle}   
    x_{\Consumer \Bundle} \leq 1,   & \Good = 1 ,\ldots, \NumberOfGoods\\\\
    
    & \displaystyle\sum\limits_{\Bundle\subseteq\SetOfGoods} 
    x_{\Consumer\Bundle} \leq 1,    & \Consumer = 1, \ldots, \NumberOfConsumers\\\\
    
    &  x_{\Consumer \Bundle} \in \{0,1\}, & \Consumer \in \SetOfConsumers, \Bundle \subseteq \SetOfGoods
\end{array}
\end{equation}

Given a market $\Market$ and a solution $\Allocation^*$ to (\ref{welfare-max-ILP}), the following set of linear inequalities, (\ref{linear-pricing-LP}), define all linear prices that couple with allocation $\Allocation^*$ to form a CE in $\Market$. 
The inequalities are defined over variables $\PriceFunction_1, \ldots, \PriceFunction_\NumberOfGoods$ where $\PriceFunction_\Good$ is good $\Good$'s price. The price of bundle $\Bundle$ is then $\sum_{\Good \in \Bundle}\PriceFunction_\Good$.

\begin{equation}
\label{linear-pricing-LP}
\def\arraystretch{0.5}
\begin{array}{lrlll}
    & 
    \Valuation_\Consumer(\Bundle) 
        - 
    \sum_{\Good \in \Bundle} \PriceFunction_\Good 
        \leq 
    \Valuation_\Consumer(\Bundle^*_\Consumer)
        -
    \sum_{\Good \in \Bundle^*_\Consumer} \PriceFunction_\Good,   & \Consumer \in \SetOfConsumers, \Bundle \subseteq \SetOfGoods\\\\
    
    & 
    \text{If } \Good \notin \cup_{\Consumer \in \SetOfConsumers} \Bundle^*_\Consumer, 
    \text{ then } \PriceFunction_\Good = 0, & \Good = 1, \ldots, \NumberOfGoods\\\\
    
    &  \PriceFunction_\Good \ge 0, & \Good \in \SetOfGoods
\end{array}
\end{equation}
\vspace{-0.15em}
The first set of inequalities of (\ref{linear-pricing-LP}) enforce the \UM{} conditions. The second set of inequalities states that the price of goods not allocated to any buyer in $\Allocation^*$ must be zero. In the case of linear pricing, this condition is equivalent to the \RM{} condition. 
In practice, a market might not have CE in linear pricings, i.e., the set of feasible solutions of (\ref{linear-pricing-LP}) might be empty. In our experiments, we solve the following linear program, (\ref{linear-pricing-LP-with-slack}), which is a relaxation of (\ref{linear-pricing-LP}). In linear program (\ref{linear-pricing-LP-with-slack}), we introduce slack variables $\alpha_{\Consumer\Bundle}$ to relax the \UM{} constraints. We define as objective function the sum of all slack variables, $\sum_{\Consumer \in \SetOfConsumers, \Bundle\subseteq\SetOfGoods}\alpha_{\Consumer\Bundle}$, which we wish to minimize. 

\vspace{-0.25em}
\begin{equation}
\label{linear-pricing-LP-with-slack}
\def\arraystretch{0.5}
\begin{array}{rllll}
\text{minimize} 
 
    \displaystyle\sum\limits_{\Consumer \in \SetOfConsumers, \Bundle\subseteq\SetOfGoods}  \alpha_{\Consumer\Bundle} \\\\
\text{subject to \quad\quad\quad\quad\quad}  \\\\

    \Valuation_\Consumer(\Bundle) 
        - 
    \sum_{\Good \in \Bundle} \PriceFunction_\Good 
        -
    \alpha_{\Consumer \Bundle}
        \leq 
    \Valuation_\Consumer(\Bundle^*_\Consumer)
        -
    \sum_{\Good \in \Bundle^*_\Consumer} \PriceFunction_\Good,   & \Consumer \in \SetOfConsumers, \Bundle \subseteq \SetOfGoods\\\\

    \text{If } \Good \notin \cup_{\Consumer \in \SetOfConsumers} \Bundle^*_\Consumer, 
    \text{ then } \PriceFunction_\Good = 0, & \Good = 1, \ldots, \NumberOfGoods\\\\
    
      \PriceFunction_\Good \ge 0, & \Good \in \SetOfGoods \\\\
      \alpha_{\Consumer\Bundle} \ge 0, & \Consumer \in \SetOfConsumers, \Bundle \subseteq \SetOfGoods
\end{array}
\end{equation}
\vspace{-0.25em}
As reported in the main paper, for each GSVM market we found that the optimal solution of (\ref{linear-pricing-LP-with-slack}) was such that $\sum_{\Consumer \in \SetOfConsumers, \Bundle\subseteq\SetOfGoods}  \alpha_{\Consumer\Bundle} = 0$, which means that an exact CE in linear prices was found. 
In contrast, for LSVM markets only 18 out of 50 markets had linear prices ($\sum_{\Consumer \in \SetOfConsumers, \Bundle\subseteq\SetOfGoods}  \alpha_{\Consumer\Bundle} = 0$) whereas 32 did not ($\sum_{\Consumer \in \SetOfConsumers, \Bundle\subseteq\SetOfGoods}  \alpha_{\Consumer\Bundle} > 0)$. 

\subsection*{Experiments' Technical Details}

We used the COIN-OR~\cite{saltzman2002coin} library, through Python's PuLP (\url{https://pypi.org/project/PuLP/}) interface, to solve all mathematical programs. We wrote all our experiments in Python, and once the double-blind review period finalizes, we will release all code publicly.
We ran our experiments in a cluster of 2 Google's GCloud \emph{c2-standard-4} machines. Unit-demand experiments took approximately two days to complete, GSVM experiments approximately four days, and LSVM experiments approximately eight days.